# WST - Widefield Spectroscopic Telescope: Motivation, science drivers and top-level requirements for a new dedicated facility.

Roland Bacon[a], Vincenzo Maineiri[b], Sofia Randich[c], Andrea Cimatti[d], Jean-Paul Kneib[e], Jarle Brinchmann[f], Richard Ellis[g], Eline Tolstoy[h], Rodolfo Smiljanic[i], Vanessa Hill[j], Richard I. Anderson[e], Paula Sanchez Saez[b], Cyrielle Opitom[k], Ian Bryson[k], Philippe Dierickx[a], Bianca Garilli[l], Oscar Gonzalez[k], Roelof de Jong[u], David Lee[k], Steffen Mieske[b], Angel Otarola[b], Pietro Schipani[m], Tony Travouillon[n], Joel Vernet[b], Julia Bryant[x], Marc Casali[o], Matthew Colless[n], Warrick Couch[p], Simon Driver[q], Adriano Fontana[r], Matthew Lehnert[a], Laura Magrini[c], Ben Montet[s], Luca Pasquini[b], Martin Roth[u], Ruben Sanchez-Janssen[k], Matthias Steinmetz[u], Laurence Tresse[t], Christophe Yeche[w], Bodo Ziegler[v]

[a]Centre de Recherche Astrophysique de Lyon ; [b]European Southern Observatory ; [c]INAF - Osservatorio Astrofisico di Arcetri ; [d]Univ. degli Studi di Bologna ; [e]Ecole Polytechnique Fédérale de Lausanne ; [f]Instituto de Astrofísica e Ciências do Espaço ; [g]Univ. College London ; [h]Kapteyn Astronnomical Institute, Univ. of Groningen ; [i]Nicolaus Copernicus Astronomical Ctr. ; [j]Observatoire de la Côte d'Azur ; [k]UK Research and Innovation ; [l]INAF - Istituto di Astrofisica Spaziale e Fisica cosmica Milano ; [m]INAF - Osservatorio Astronomico di Capodimonte ; [n]The Australian National Univ. ; [o]Macquarie Univ. ; [p]Swinburne Univ. of Technology ; [q]The Univ. of Western Australia ; [r]INAF - Osservatorio Astronomico di Roma ; [s]The Univ. of New South Wales ; [t]Lab. d'Astrophysique de Marseille; [u]Leibniz-Institut für Astrophysik Potsdam ; [v]Univ. Wien ; [w]CEA IRFU; [x]Univ. of Sydney

## ABSTRACT

In this paper, we describe the wide-field spectroscopic survey telescope (WST) project. WST is a 12-metre wide-field spectroscopic survey telescope with simultaneous operation of a large field-of-view (3 sq. degree), high-multiplex (20,000) multi-object spectrograph (MOS), with both a low and high-resolution modes, and a giant 3×3 arcmin$^2$ integral field spectrograph (IFS). In scientific capability, these specifications place WST far ahead of existing and planned facilities. In only 5 years of operation, the MOS would target 250 million galaxies and 25 million stars at low spectral resolution, plus 2 million stars at high resolution. Without need for pre-imaged targets, the IFS would deliver 4 billion spectra offering many serendipitous discoveries. Given the current investment in deep imaging surveys and noting the diagnostic power of spectroscopy, WST will fill a crucial gap in astronomical capability and work in synergy with future ground and space-based facilities. We show how it can address outstanding scientific questions in the areas of cosmology; galaxy assembly, evolution, and enrichment, including our own Milky Way; the origin of stars and planets; and time domain and multi-messenger astrophysics. WST's uniquely rich dataset may yield unforeseen discoveries in many of these areas. The telescope and instruments are designed as an integrated system and will mostly use existing technology, with the aim to minimise the carbon footprint and environmental impact. We will propose WST as the next European Southern Observatory (ESO) project after completion of the 39-metre ELT.

# 1. INTRODUCTION

Astrophysics is experiencing a golden era with many breakthrough discoveries and a variety of new instrumentation and programmes planned for the next decade to pursue these discoveries. Nonetheless, a comprehensive understanding of the formation and evolution of structures in the Universe is still missing. Many key questions in contemporary astrophysics and cosmology remain unadressed and can only be resolved with new observational data:

- Is the accelerated expansion of the Universe due to an unknown form of energy or to a modification of General Relativity on large scales?
- What is the interplay between dark, stellar, and gaseous material in galaxies and how does primordial and metal-enriched gas flow in and out of galaxies at various scales?
- What is the detailed formation history of our own Galaxy, the Milky Way and of its satellites?
- What is the origin of the chemical elements that are crucial to trace galactic evolution?
- What are the conditions that drive the formation and evolution of extra-solar planets?
- What are the extreme physical conditions that govern transient events (explosions, eruptions, and disruptions)?
- What do gravitational waves tell us about neutron star physics, heavy element production, and cosmology?

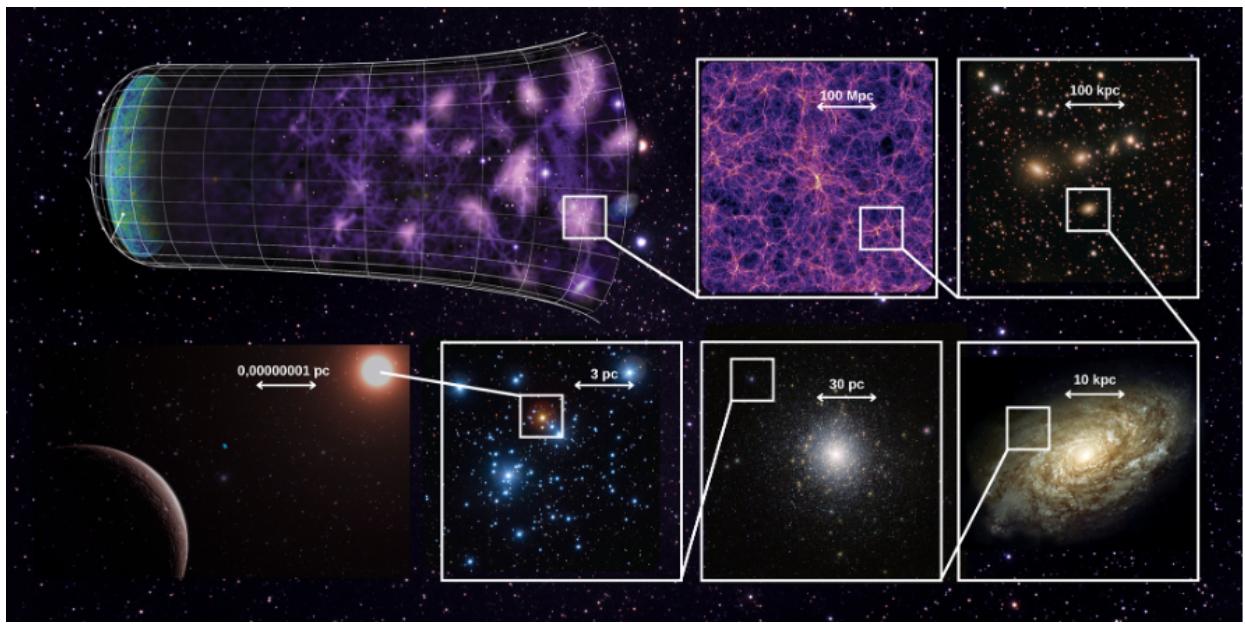

*Figure 1: WST will be a revolutionary spectroscopic facility addressing many open questions in astrophysics over a large range in physical scales: from the formation of the large-scale structures in the early universe (100s of Mpc), to the interplay of galaxies in the cosmic web (10s of Mpc), to the formation of our own Galaxy (kpc scales), to the evolution of stars and the formation of planets around them (sub-pc scales). Credits: Rossella Spiga - INAF – OAArcetri.*

A crucial element in answering these questions will be the availability of vast spectroscopic datasets covering a large fraction of the sky.

Over the next decade, we expect a deluge of high-quality imaging data to be delivered by the upcoming ground-based — e.g., Vera Rubin Observatory (LSST), SKAO, CTA, Einstein Telescope (ET) — and space-based — e.g., JWST, Euclid, Roman Space Telescope, LISA — survey-oriented facilities. In parallel the ESA *Gaia* astrometric mission is expected to have its final data release in 2030. These new facilities will detect and image an enormous number of celestial objects with unprecedented precision. However, to fully characterise and understand them requires spectroscopic follow-up at adequate spectral resolution and cadence, both for representative sources and rarer phenomena, the latter of which can lead to exciting new discoveries. Given the expected number of sources (e.g., 20 billion galaxies and 17 billion stars down to R~27.5 from LSST alone), only a dedicated wide-field spectroscopic facility will fully realise the scientific potential of the huge investment made in these imaging surveys. Such a partnership between survey imaging and follow-up spectroscopy has been ably demonstrated by many earlier stellar and extragalactic surveys[1] that have had a lasting scientific impact and strong legacy value.

Multi-object spectroscopic (MOS) surveys are usually performed with instruments based on fibre optic (e.g., FLAMES at VLT) or multi-slit technologies (e.g., VIMOS at VLT). Recent MOS facilities using dedicated 4m-class telescopes are operational or coming on-line shortly (e.g., DESI, 4MOST, WEAVE). In parallel, new MOS instruments will soon be installed on existing 8-10m class multi-purpose telescopes (e.g., PFS at Subaru, MOONS at VLT). However, these new facilities will only perform surveys of relatively bright sources (in the case of dedicated 4m facilities) or limited campaigns of faint sources (for multi-purpose 8-10m telescopes). Upcoming giant telescopes (ELT, TMT, GMT) are also limited, by optical practicality and technical feasibility, to a small field-of-view and a correspondingly modest multiplex capability (e.g., 300 over 0.12 degree$^2$ for MOSAIC at the ELT). In addition, these are multi-purpose telescopes for which only a small fraction of the time would be available for this survey science. Consequently, only a dedicated wide field 10m-class telescope equipped with a very high-multiplex MOS will be able to fully exploit the upcoming multi-wavelength imaging data, as well as efficiently follow-up the wide range of transients and time-variable phenomena revealed by LSST.

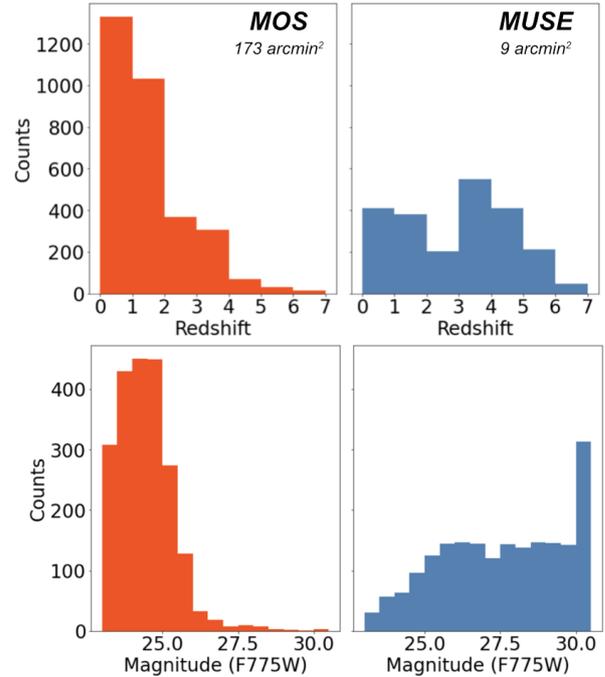

*Figure 2: Example of complementarity between extragalactic MOS and IFS spectroscopic surveys. The left panels present the redshift (top) and magnitude (bottom) distributions of all spectroscopic redshifts obtained from a decade of efforts with 10m-class MOS instruments in the CDFS wide field (~173 arcmin$^2$). The right panels show those recently measured by MUSE in the small HUDF field (8.2 arcmin$^2$) within the CDFS area (mosaic of 3x3 MUSE exposures) [1].*

Meanwhile, the advent of panoramic integral-field spectrographs (IFS) on 8m-class telescopes (e.g., MUSE) has been made possible by well-established IFS expertise in Europe, and in particular technological advances in large format glass slicers. Their field of view (e.g., 1 arcmin$^2$ for MUSE) and exquisite sensitivity has enabled very deep spectroscopic surveys in smaller patches of the sky. Collectively, the MOS and IFS spectroscopic approaches are highly complementary: the MOS surveys large samples over wide areas, whereas the IFS studies densely packed fainter sources on finer scales as illustrated in Figure 2. As an example, a MOS can detect galaxies to a magnitude 25 (median) and redshift $z < 3$ over representative cosmic volumes, while an IFS can detect a similar number to much fainter magnitudes (median AB ~27) and at higher redshifts ($z < 7$) over a smaller area. The combination allows both bright and faint ends of the galaxy luminosity function to be probed over a wide redshift range (z=0-7).

A panoramic IFS has a further unique property. Unlike the MOS, as it samples all sources within its field of view, no target pre-selection is required. The IFS offers 2D continuous sampling enabling the discovery of faint extended structures (e.g., emission from circumgalactic/intergalactic gas) as well as previously unknown sources. The main limitation of current IFS is their relatively modest field-of-view. A larger field-of-view is highly desirable, but given the very large number of spectrographs involved (144 in this proposal), it is not compatible with the infrastructures on current 10m-class telescopes[2]. Clearly, a new dedicated, purpose-built facility is the only way forward.

---

[1] For example: 2dF, SDSS, APOGEE, GALAH, Gaia-ESO, RAVE, GAMA, VVDS, VIPERS
[2] As illustrated in Figure 13, where the huge volume required by the 144 IFS spectrographs can only be accommodated by a specific opto-mechanical design of the facility.

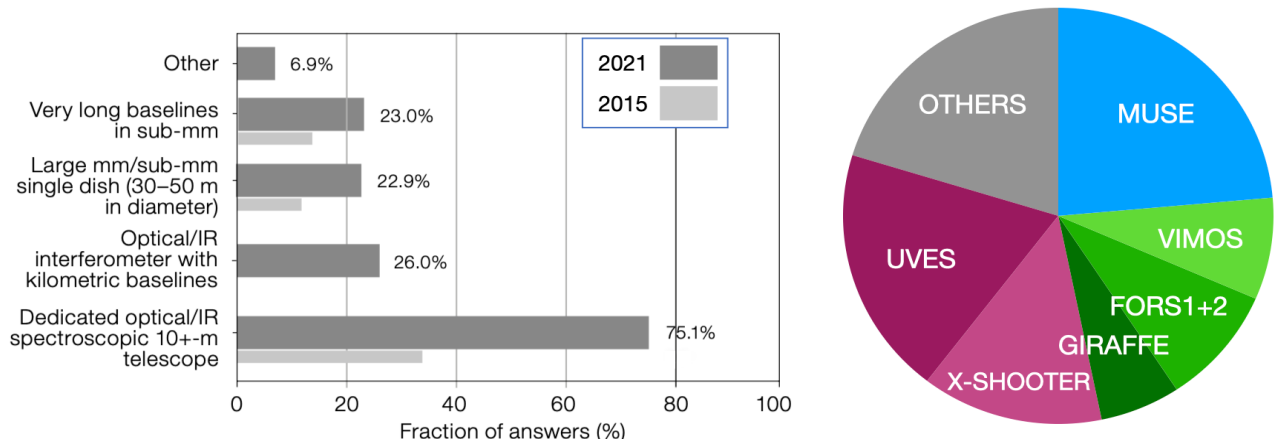

*Figure 3 **Left**: Importance of present and future (planned or potential) facilities from "Report on the ESO scientific prioritization community poll" [2]. The thinner and lighter grey bars show the opinions from the 2015 survey (where available). **Right**: Statistics of 720 refereed papers published in 2021 for all VLT instruments. The papers produced with the panoramic IFS (23%, blue sector), low and medium resolution MOS (22%, green sectors) and high-resolution single spectrographs (32%, magenta-violet sectors) represent the vast majority (77%) of all publications using VLT instruments, leaving less than one third to those produced by all other VLT instruments. Source ESO Telbib.*

The scientific demand for a 10m-class wide-field telescope dedicated to spectroscopic surveys is internationally recognised since it features prominently in many national strategic plans (e.g., US 2020 decadal survey, 2016-2025 decadal plan for Australian astronomy, Canadian astronomy long range plan 2020-2030). In Europe in particular, a recent poll among ESO users demonstrated that 75% of that community support such a facility as the most crucial one for the future (Figure 3, left panel). The concept of combining both MOS and IFS capabilities in a dedicated wide-field 10m-class spectroscopic telescope would ensure full exploitation of the imaging data from upcoming facilities. A spectroscopic facility will have a very broad user community as reflected by the scientific productivity of current VLT spectroscopic instruments (Figure 3, right panel). Independent analysis has led Astronet - the planning and advisory network for European astronomy - to prioritise a general-purpose wide-field high-multiplex spectroscopy facility as one of the recommended major ground-based infrastructure developments. The other two are the Cherenkov Telescope Array (CTA) and the European Solar Telescope (EST).

Such a facility will enable the scientific community to comprehensively address the questions outlined above and lead to transformative science in several key areas, from cosmology to the formation of stars and planets. Crucially, as with all facilities with unique capabilities, it will lead to unforeseen discoveries of great significance.

## 2. TOP LEVEL REQUIREMENTS

*Table 1: WST top-level requirements*

| Telescope Aperture | 12 m, seeing limited |
|---|---|
| Telescope FoV | 3.1 deg² |
| Tel. Spec Range | 350 -1600 nm |
| MOS LR Multiplex | 20,000 |
| MOS LR Resolution | 3,000-4,000 |
| MOS LR Spec Range | 370-970 nm (simultaneous) |
| MOS HR Multiplex | 2,000 |
| MOS HR Resolution | 40,000 |
| MOS HR Spec Range | 370-970 nm (3-4 regions) |
| IFS FoV | 3x3 arcmin² |
| IFS Resolution | 3,500 |
| IFS Spec Range | 370-970 nm (simultaneous) |
| IFS Patrol Field | 13 arcmin (diameter) |
| MOS & IFS simultaneous operation | |
| ToO implemented at telescope and fibre level | |

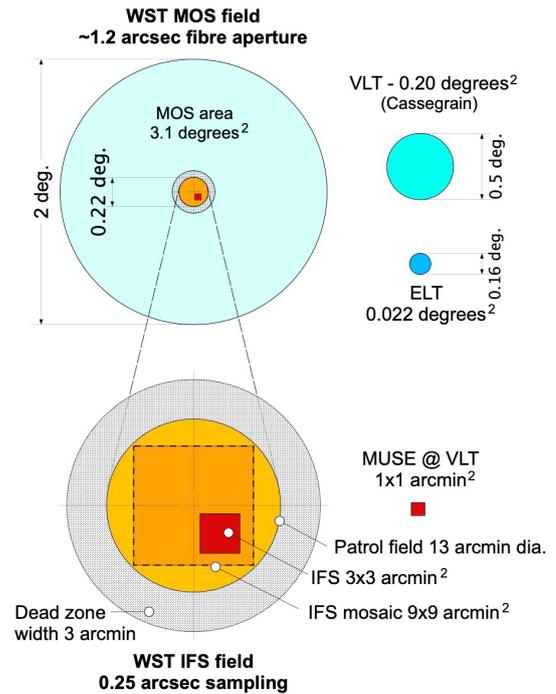

*Figure 4: WST fields of view (FoV). The top panel shows the MOS FoV and the central circular area available for IFS observations. The bottom panel offers a closer view of the latter. The IFS 3x3 arcmin² FoV (in red) can be moved within the available area, providing the 9x9 arcmin² mosaic capability A distinct dead zone separates the MOS and IFS FoVs. The VLT, ELT and MUSE FoV are represented for* comparison.

We are currently working on a concept study for a 10m-class wide-field spectroscopic survey telescope with simultaneous operation of a large field-of-view, high-multiplex MOS, having both low- and high-resolution modes, and a giant panoramic central IFS. From the science cases devised by our international science team we have defined the set of preliminary top-level requirements set out in Table 1. WST is designed as a 12m aperture active[3] telescope that covers the UV to the H band and operates in seeing limited conditions. The field-of-view (FoV) of 3.1 deg² will be the largest of any 8m-class telescope equipped with a MOS and, in particular, its FoV will be 2.4 times larger than the Japanese PFS@Subaru and a remarkable 22 times larger than MOONS at ESO VLT (*Figure 4*). In addition, the WST has a substantial aperture, twice as large (100 square metres) as its 8m class competitors. A central circular area with a diameter of 13 arcmin is available for simultaneous IFS observations. Within this area, the IFS with a FoV of 3´3 arcmin² (9 times larger than MUSE) can be utilised simultaneously with the MOS. In addition, the IFS can be used to mosaic a larger area within the 13 arcmin circle (an area larger than the entire FoV of the ELT) while the MOS positioner can perform multiple configurations (*Figure 4*). The WST operational model will also implement Target of Opportunity (ToO) operations at both the telescope and MOS fibre levels to support the time domain science cases.

These ambitious requirements will place the WST MOS and IFS capabilities far ahead of the existing or planned competing facilities (Figure 5). Using the collecting area and the fibre multiplex gain as a proxy for the survey speed, WST MOS will be faster by x 18 than PFS@Subaru and by x 3.4 than the proposed US Megamapper facility. We note that while MSE-QM, the latest proposed quad-mirror version of the Mauna Kea Spectroscopic Explorer (MSE), achieves similar MOS performance using this metric, the larger field of view of WST gives it a factor of 2 better etendue[4]. The WST IFS, with its large field of view, good spatial resolution and large simultaneous spectral range, will outperform all existing and planned panoramic IFS. For example, using the etendue metric, it will be 19 and 162 times faster than the European MUSE@VLT and the US KCWI@Keck, respectively.

---
[3] The implementation of adaptive optics for the MOS system is not realistic due to the large FoV and short wavelengths. Even the IFS patrol area, which has a much smaller FoV, is unlikely to benefit significantly from a GLAO system. In addition, the integration of such a system would add significant complexity and cost.

[4] Etendue is the field-of-view area times the primary mirror collecting area.

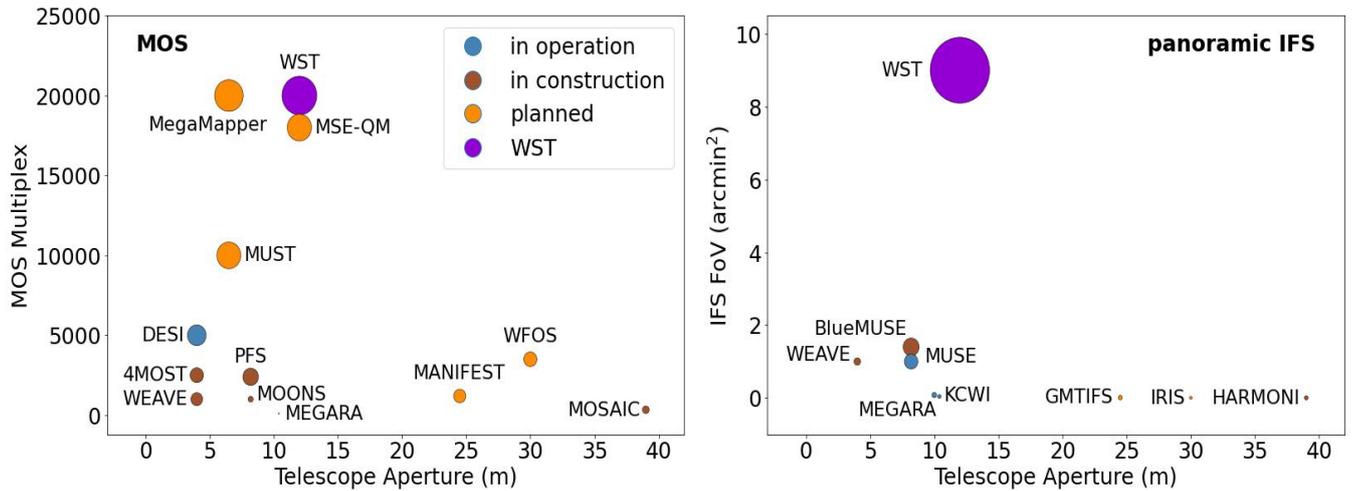

*Figure 5: Comparison of WST MOS (left panel) and IFS (right panel) capabilities with existing and proposed ground-based spectroscopic facilities. Circle areas are proportional to the etendue (i.e., aperture times field of view area). For clarity, MOS or IFS with small multiplex or field of view are not shown in these figures. Multi-IFUs (e.g. KMOS, Hector or MaNGA) or sparse-field panoramic IFS (e.g. HETDEX) are not shown in this comparison, which only considers panoramic (monolithic) IFS with 100% fill-factor. However, multi-IFUs are considered as a possible WST upgrade to the MOS components.*

It should be stressed that neither Megamapper nor MSE have IFS capabilities. Thus, not only will WST perform significantly better than any existing or planned MOS or IFS instruments, but the simultaneous operation of its MOS and IFS capabilities as a single facility will represent a significant advance in the overall scientific capability available to astrophysicists and cosmologists.

WST will thereby provide more science output for a given amount of telescope time and offer flexibility in the survey design. For example, it will enable the design of a MOS-driven survey of the large-scale distribution of galaxies while, at the same time, on-axis IFS deep field exposures to measure the properties of high redshift line emitting galaxies. Likewise, an IFS-driven survey of the core of galaxy clusters can be undertaken while using the large field MOS to study galaxy properties in the outskirts of those clusters.

The initial set of instruments is constrained to the UV-visible range (370-970 nm), for two reasons. The first is cost, as hundreds of IR detectors would, at current prices, be prohibitively expensive. The second is risk, as accurate sky subtraction of fibre-fed spectrographs in the IR and at faint limits is yet to be demonstrated. To allow future upgrades, which future technology and instruments development may plausibly allow, and considering the expected lifetime of the facility, the telescope has been designed to cover the UV to H-band (350-1600 nm). Such upgrades would enhance the science capabilities of surveys and ensure WST' competitiveness over an extended lifetime. Beside an extension into the near-infrared, upgrades may also consider for example the deployment of multiple IFUs or the addition of an intermediate spectral resolution (R~10,000).

The simultaneous integration of IFS and MOS capabilities is a logical next step in terms of scientific capabilities but unique in the current landscape of spectroscopic facilities. While offering remarkable flexibility and observational efficiency, it requires an entirely new operational model. The massive MOS multiplex capability enables the execution of multiple, scientifically distinct, surveys concurrently. The need for a rapid response to time domain discoveries contributes further to the complexity of observing programmes. This implies the challenging task of developing an operational model which allows "focal plane sharing" between the two different instruments, and between surveys with different scientific goals. The experience gained with 4MOST, which will be operating in a similar way, will be valuable in addressing WST's operation challenges.

Over its lifetime, with such a large field and a huge multiplex capability, WST will provide massive spectroscopic data sets. Timely delivery to the community for full and rapid scientific exploitation becomes fundamental. This demands the development of efficient tools for processing raw data into science-ready products, as well as for archiving and data mining. The needs or new algorithms based on state-of-the-art information must be part of the evaluation study, and will have to be implemented by the start of WST operations.

Along with the facility concept study and the big data challenges, sustainability also represents a crucial aspect. Hence, we will emphasise flexibility in the design which will make future upgrades and extensions to the telescope, ensuring a long effective lifetime. Today, at a time of climate crises, we must be cognisant of the environmental impact of astronomical infrastructures. We plan to develop WST as model facility in this regard by minimising the $CO_2$ footprint and environmental impact both during construction and operations. For the site selection, we will consider the ability of the facility to rely on solar and other renewable energy sources, such as the 9MW solar array that serves the Paranal Observatory in Chile (Figure 14). The cooling processes for the detectors will be developed to minimise the energy required to effectively maintain their operating temperatures. Operations will be structured to minimise the effects on both the local environment and greenhouse gas emissions, for example by reducing waste and water usage to the greatest extent possible. When the telescope is operational, data processing will aim to use storage and supercomputing resources that are sustainable.

## 3. SCIENCE CASE

Due to its unique capabilities, WST will provide an unprecedented and panoramic view of the formation of large-scale structures and galaxies over most of cosmic history, unveil the formation and evolution of our own Galaxy and its satellites, and reveal the extreme physical conditions producing spectacular astrophysical phenomena such as supernovae, and tidal disruption and gravitational wave events. The scientific impact of WST will be transformative in each of these different areas.

In its first 5 years of operation, we estimate WST will provide:
**MOS MR**: **250 million galaxies** (to mag 24.5) over 18,000 deg$^2$ (the entire accessible sky from north Chile, excluding the Galaxy) and **25 million stars** (to mag 23.0) over the entire Galaxy and the Local Group visible from north Chile, with spectral measurements of redshifts, galaxy kinematics, stellar ages and metallicities, etc.
**MOS HR**: **2 million stars** (to mag 17.0) over the entire Galaxy and the Local Group visible from north Chile, with spectral measurements of chemical abundances, stellar kinematics, etc.
**IFS**: **4 billion spectra** over 30 deg$^2$ in diverse environments (low-density fields, galaxy and star clusters, Galactic fields …) with spatially-resolved spectral properties and redshifts, intergalactic- and circumgalactic-medium properties, etc.

The box at right outlines the projected numbers of sources or spectra that WST will observe within the first 5 years of operation. To put these figures in context, reproducing the 4 billion spectra generated by the WST IFS would necessitate 43 years of VLT time fully dedicated to MUSE, while conducting a survey of 250 million galaxies at the same depth as WST with 4MOST would take 375 years.

The widespread scientific interest in WST is evidenced by a Science Team that includes more than 500 researchers from 32 countries across five continents. The Science Team has recently published a comprehensive description of the transformational science cases that this new facility will tackle in the form of a **WST Science White Paper** [3]. Referring the reader to the White Paper for a comprehensive description of WST's scientific potential, here we provide some illustrative examples highlighting the expected scientific impact of WST.

*The nature of dark energy and gravity in the early Universe.* Understanding the nature of the dark Universe and its accelerated expansion is one of the key questions in fundamental physics and cosmology. The large-scale structure (LSS) of the Universe provides powerful tools for addressing this question because it encodes information that can reveal whether this acceleration is due to an unknown form of energy or requires a modification of General Relativity at large scales. Measurements of galaxy clustering constrain the nature of dark energy via the scale of baryon acoustic oscillations (BAO, pressure waves imprinted on the early universe) and gravity via the growth rate of structure (obtained from redshift space distortions induced by the peculiar velocities of galaxies).

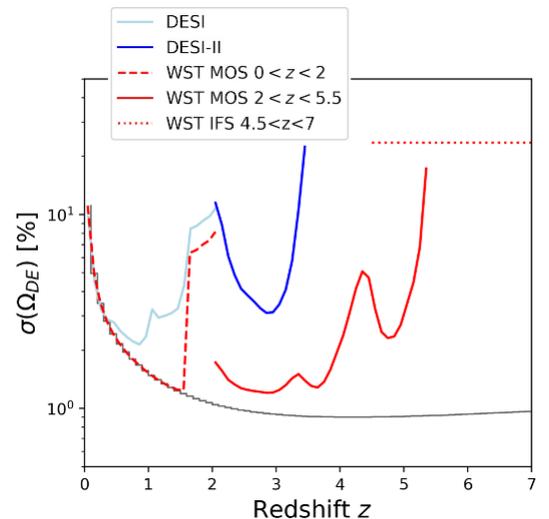

*Figure 6: WST (red curves) will allow to constraint the dark energy models to a level of precision unachievable from current or upcoming facilities (DESI, DESI-II) and open a completely unexplored parameter space at z>4.*

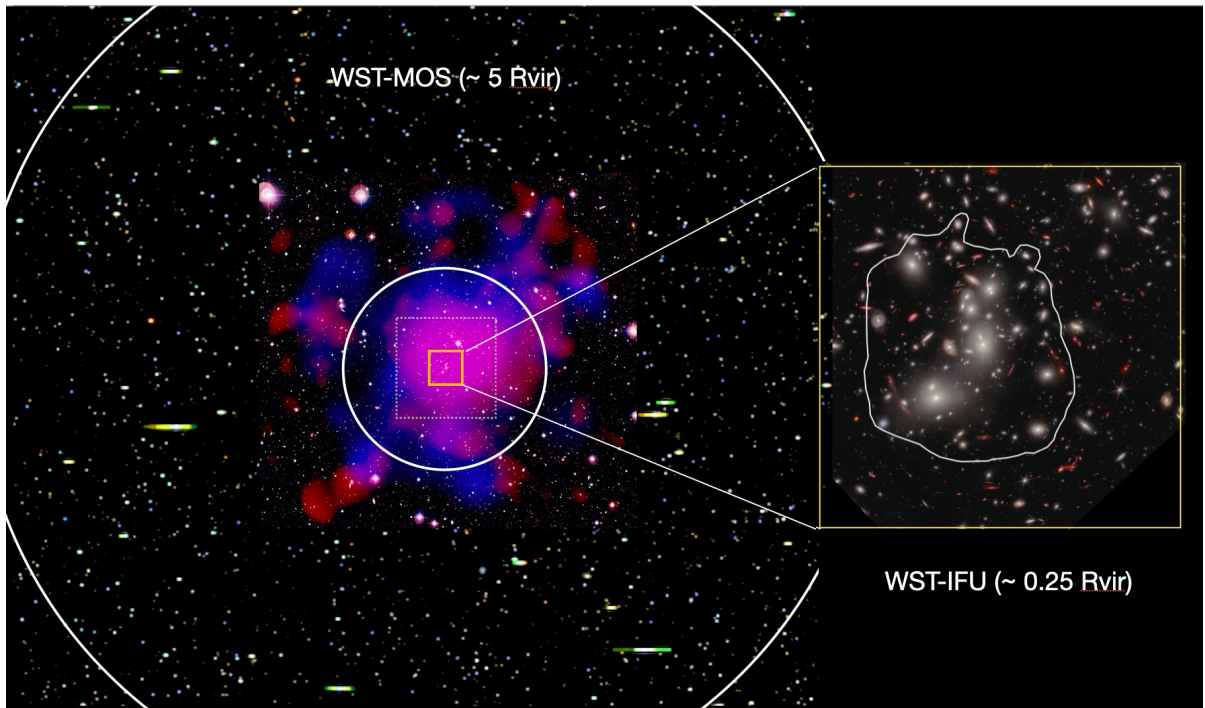

*Figure 7: A schematic of the WST comprehensive view using MOS and IFU of the massive clusters in the Universe - measure of their mass (using lensing, X-ray and dynamics), how they grow with time and how they connect to the filaments and large-scale structures. .Credits: Johan Richard.*

While the current generation of redshift surveys exploiting telescopes with apertures up to 4 metres (e.g. DESI, Euclid, 4MOST) are probing the universe up to redshift $z$=2, a WST cosmology survey over a footprint of 18,000 deg$^2$ will be able to constrain the scale of the BAO and the growth rate of structure with unprecedented precision at redshifts inaccessible to current facilities ($2 < z < 5.5$). Selecting targets galaxies through different techniques (e.g. Lyman-break galaxies, LBG), WST will be able to probe these high redshifts, when (in contrast to today) the Universe is dominated by matter, the expansion is decelerating, and structures grow with high efficiency. Such an ambitious survey will allow us to conduct sub-percent galaxy clustering measurements and consequently constrain fundamental cosmological constituents and forces in the early Universe with unprecedented sensitivity (Figure 6).

The cosmological model will ultimately also determine the properties of the most massive structures in the Universe: galaxy clusters. Galaxy clusters provide a privileged laboratory bridging cosmology and galaxy formation. As the latest virialised structures forming around the densest nodes of the cosmic web, they keep a memory of the primordial density field while hosting galaxies at the latest evolutionary stage. WST will be able to probe an unprecedented statistical sample of ~7000 clusters at redshifts $0.1<z<1.5$ by fully exploiting its unique capabilities: exploring the inner regions with the IFS and the outskirts and connections to the cosmic web of large-scale structure with the MOS (Figure 7). The 3D galaxy density field on the periphery of the cluster can be mapped by MOS measurements of the positions and redshift of galaxies, from which their orbits and the velocity field can be reconstructed. Simultaneously, the IFS can chart the position and 2D kinematics of galaxies. By synthesising the dynamics of the whole system and inferring the past orbits of cluster galaxies, the role of the environment in shaping galaxy evolution can be assessed and the physical properties of galaxies in dense environments (metallicity, SFR, etc.) related to the types of environment (voids, sheets, filaments) they passed through on their journey into the cluster.

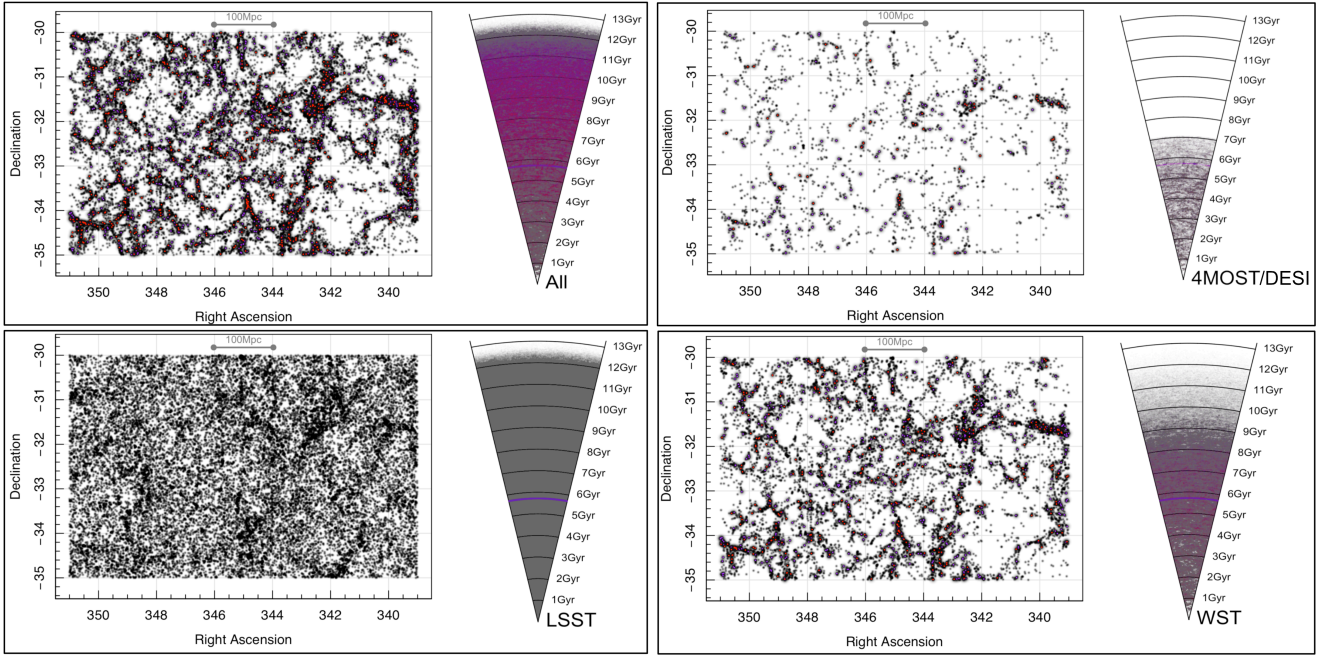

*Figure 8: The 3D dark-matter distribution of the Universe is revealed by the galaxy distribution. Here we show a narrow slice at 6 Gyr taken from the observation cone. The upper left panel shows the actual dark matter distribution, while the upper right and lower left panels present the dark matter distribution as they will be observed by 4MOST/DESI and LSST, respectively. The lower right panel displays the dark matter distribution revealed by a WST baseline (<5yr) survey.*

*Dark matter mapping with WST:* WST will directly map the dark matter distribution out to redshift $z$=1. The dark matter distribution defines the underlying substructure of our Universe. It is responsible for building the large scale-structures that we see in the form of interlocking voids, sheets, filaments, groups and clusters. Different flavours of dark matter (cold, warm, fuzzy, self-interacting, etc.) predict differences in the detailed structure of the dark matter distribution, which surveys with WST can distinguish.

Contemporary numerical simulations predict strong growth and evolution of the dark matter distribution: from a near-uniform 3D distribution at high redshift, to flattened interlocking 2D sheets at intermediate redshift, and 1D intertwining tendrils and growing voids at low-redshift (see Figure 8, top left). This structure and topology evolution is best revealed by a high-completeness spectroscopic mapping of the galaxy distribution, the identification of the dark matter concentrations (clusters and groups), and dynamical mass measurements from their velocity dispersions.

About half the dark matter in the present-day ($z$=0) Universe is expected to reside in haloes of Milky Way mass and above. A baseline (<5yr) survey with WST will identify all clustered structures down to Milky Way halo masses at intermediate redshift, yielding a direct map of the dark matter distribution (Figure 8, bottom right) and its evolution over the past 8 billion years. High-order statistics of the observed distribution can be matched with bespoke numerical simulations to distinguish flavours of dark matter and the anticipated evolution of their structure/topology. Moreover, a galaxy's evolution is likely regulated strongly by the halo in which it resides, opening a door to study galaxy populations as a function of the underlying dark matter halo mass (currently lacking beyond $z$=0.2).

In the coming decade, facilities such as DESI and 4MOST will take the first steps in this direction (Figure 8, top right), but ultimately the 12m aperture, wide-area capability and high multiplexing of WST will be critical to stringent tests of different dark matter scenarios. While robust photometric redshifts from facilities such as LSST, or weak lensing studies from Euclid/Roman, can statistically recover sub-structure, such maps are coarse (Figure 8, bottom left) and individual halo masses poorly constrained. A WST survey of ~20 million galaxies covering several hundred square degrees, possible in the first 5 years, will directly reveal over half the dark matter content of a cosmologically representative volume out to $z$=1. This will provide an unprecedented test of contemporary dark matter paradigms.

*The small-scale matter cycle.* Galaxy evolution is driven by feedback processes occurring on a variety of scales. On the scales of giant molecular clouds, the star formation efficiency and multi-phase interstellar medium (ISM) is shaped by small-scale processes. On galaxy-wide scales, outflows and secular evolution drive the interplay between galaxies, their dark matter halos, and the circum- and intergalactic media. Understanding the distribution of baryonic matter and energy across these scales remains a key challenge for both theoretical and observational astrophysics. Observationally, large galaxy surveys provide the statistics needed to study the assembly of galaxy subcomponents, and the link between galaxies and their large-scale environments. Addressing the physics of star formation, however, requires a statistical perspective on the individual components of the matter cycle (HII regions, star clusters) at a spatial resolution comparable with the size and separation length characteristic of star-forming regions (~100 pc).

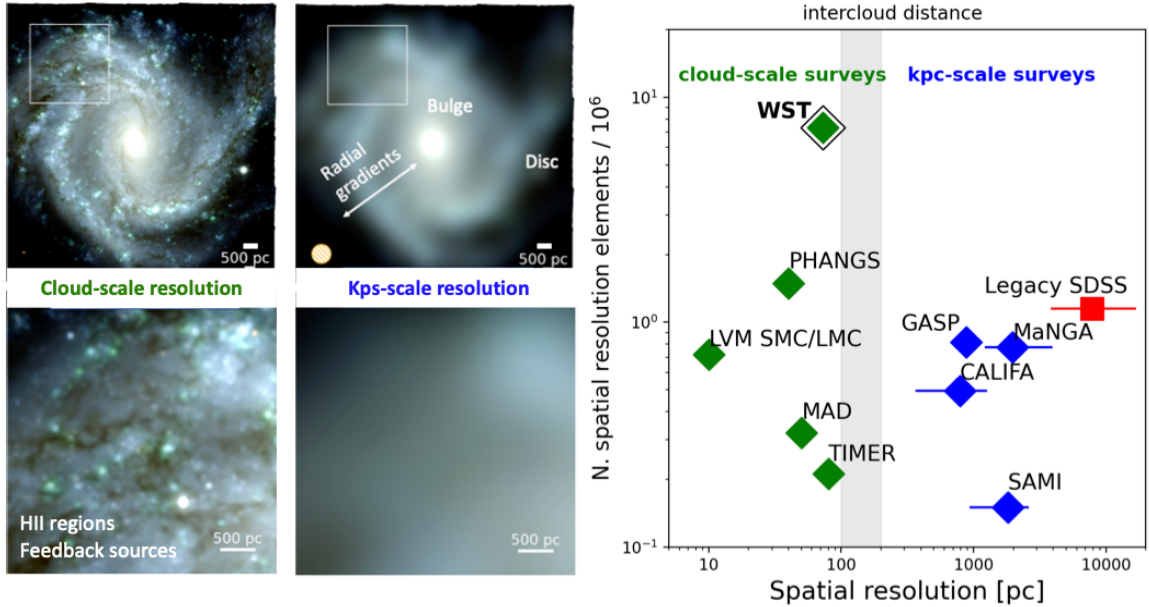

*Figure 9. Left: an illustration of the level of detail achievable in surveys that resolve the average distance between HII regions ('cloud-scale') and those that observe galaxies at kpc resolution. Right: comparison of a WST survey of the Local Volume with existing cloud-scale and kpc-scale surveys. The grey band indicates the typical distance between clouds. Only WST will be able to provide a statistical sample of galaxies at cloud-scale resolution. Adapted from Emsellem et al. (2022).*

The current generation of IFS (e.g., MUSE at the VLT) has enabled the mapping of nearby galaxies, demonstrating the power of this method. However, such samples are relatively modest and span a limited range of environments within galaxies (typically the inner few kpc), impeding our ability to match any insights drawn with results obtained from much larger galaxy surveys probing kpc scales (e.g., MaNGA and SAMI). To make progress requires 'cloud-scale' (<100 pc) mapping of a sufficiently large sample of galaxies ($>10^3$) across a range of morphological features (e.g., spiral arms, bars, bulges) and integrated properties (e.g., $M_{star}$, SFR, local environmental density). With its large IFS, WST is uniquely placed to perform an ambitious spectroscopic survey in the Local Volume (D < 25 Mpc) at better than 100 pc resolution (Figure 9) to address the challenging question of the small-scale matter cycle. Resolving individual nebulae will allow statistical studies of the metallicity of HII regions, in combination with the stellar population, characterizing the small-scale production and flow of metals in galaxies. WST's large aperture will ensure the detection of weak auroral lines yielding direct metallicity estimates beyond the inner gas-rich regions in poorly explored environments for star formation, where conversion from neutral to molecular hydrogen becomes highly inefficient. A sample of ~$10^4$ regions would adequately span approximately 2 dex in the metallicity, N/O ratio, and ionization parameter. Such a sample would be two orders of magnitude larger than the current state-of-the-art (e.g., CHAOS). Gas phase metallicity measurements will be combined with those from integrated-light spectroscopy and, (where available) resolved to chart the local history of chemical enrichment, the impact of metal diffusion, and the role of morphological features (e.g. spiral arms, bars) in mixing.

*The formation history of the Milky Way from chemical tagging and chemical clocks.* The stars in the Milky Way and its satellites contain detailed information about Galaxy formation and evolution; indeed, stars record the past in their ages, photospheric chemical compositions, and kinematics. A uniquely powerful approach to understanding the details of stellar evolution, planet formation, and galaxy formation and evolution is to measure the elemental abundances in the photospheres of large numbers of stars over the full age range, with and without known planets. Buried in these stars lies the historical record of past events throughout the large volume in which these stars live.

Stars migrate across the Milky Way, eventually reaching regions that are very far and different from those where they formed. The time variation of the Galactic potential and of transient structures such as the bar and spiral arms precludes the use of kinematics alone in tracing stars back to their birth position. In contrast, elemental abundances provide unique fingerprints that identify the environment in which stars formed. This means that abundances can be used to tag stars that form within a stellar association that has a unique and recognizable pattern (so-called 'chemical tagging'). The main requirement to make a fundamental breakthrough in chemical tagging is to obtain precise and accurate abundances with errors less than 0.05 dex. This required level of precision can only be achieved with spectra with high signal-to-noise ratios (SNR>100) and high spectral resolution (R~40,000). The latter is essential to have enough unblended spectral lines for precise abundance determinations (Figure 10). WST's unique combination of high multiplexity, high resolution mode, and large aperture will allow us to observe all main-sequence turn-off stars up to distances from the Sun of 3 kpc and all intrinsically brighter red giant stars up to ~10-30 kpc with the required SNR. WST will increase by at least a factor of ~30 the volume accessible to this kind of study compared to current or upcoming facilities (Figure 11).

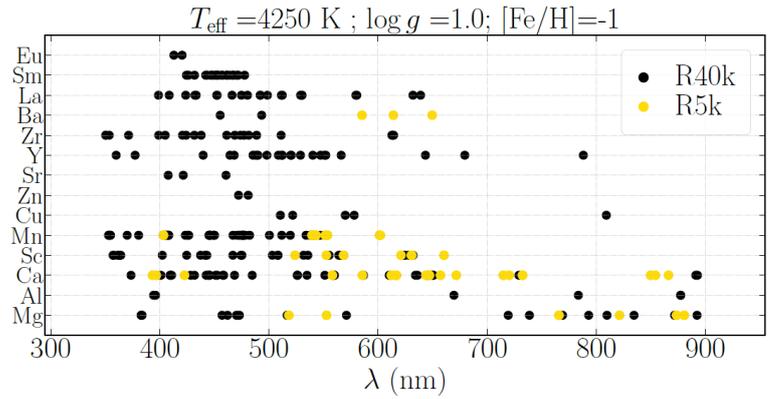

*Figure 10: The line-detectability for a few cherry-picked elements (listed on the y-axis). High-spectral resolution is essential to dramatically improve the number of suitable spectral lines: more than a factor of 5 comparing R=40K (black points) with R=5K (yellow points). Adapted from [4].*

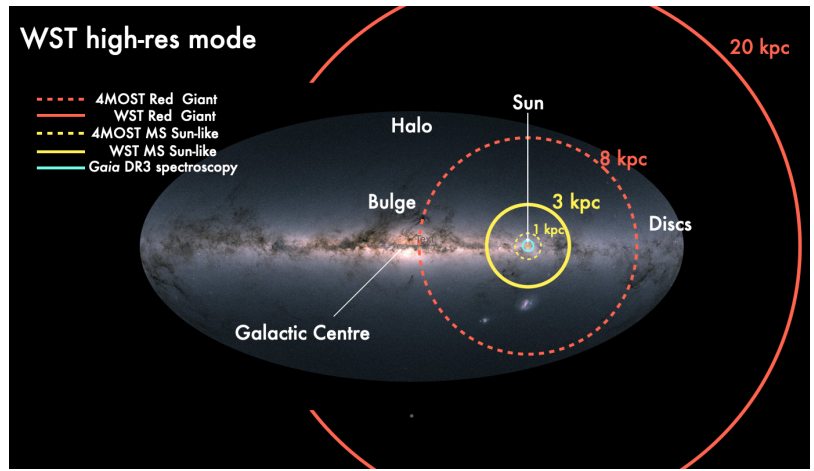

*Figure 11. Sky map with circles comparing the regions up to where a certain stellar type can be observed in high-resolution mode with WST (red and yellow solid lines) and 4MOST (red and yellow dashed lines). The region where abundances from Gaia DR3 spectroscopy is available is shown in cyan.*

WST will also greatly enhance our ability to measure abundance ratios that are sensitive to age (so-called 'chemical clocks', which involve abundances of neutron capture elements and light elements such as C and N). The exploitation of these clocks requires careful calibration, based on samples of stars with reliable age determinations, such as open clusters or asteroseismic targets for which asteroseismology information is available. Well-calibrated chemical clocks can be used to produce extensive age maps of the disc, crucially expanding our understanding of the various stellar populations and how they have moved with time. In a 5-year survey with WST, it would be possible to obtain SNR>100 spectra for $10^7$ stars, enabling the detection of about $10^4$ fossil groups/clusters, now disrupted, and thus leading to a revolutionary advance in this important field.

*Electromagnetic counterparts of gravitational wave events.* Time-domain and multi-messenger astronomy are fundamental priorities identified by major planning exercises, such as the Astro2020 decadal survey, and they are tightly connected since multi-messenger events are typically of a transient nature. Massive investments in large spectroscopic facilities are required to ensure accurate classification and characterization of variable and transient objects identified by large surveys (e.g., *Gaia*, LSST). Unfortunately, all current and upcoming facilities together can follow up only a few percent of the expected transient alerts. WST is being conceived with time-domain science as one of its core priorities. To this end, WST will provide maximum operational flexibility and rapid data processing to serve the needs of this exciting and diverse field full of serendipitous discoveries, with science cases spanning all physical scales from Solar system objects to Cosmology and all-time scales from hours to decades. WST will make game-changing contributions to the multi-messenger revolution which was heralded by detections of gravitational wave (GW) emitters with electromagnetic counterparts. WST will provide much-needed follow-up and characterization observations of sources discovered by next-generation detectors, such as the Einstein Telescope (ET) and Cosmic Explorer (CE). ET will detect ~$10^5$ GW events from merging binary neutron stars (BNS). ET will even discover some BNS during the inspiral phase up to hours before the merger. WST's simultaneous IFS & MOS operations will be perfectly suited for providing the required spectroscopic characterization at early times thanks to telescope-level target of opportunity observations as illustrated in Figure 12. The IFS would survey high-density and high-probability regions, while the MOS fibres would be placed on live electromagnetic events within the FoV. The IFS will be particularly useful to capture nearby BNS events that can occur significantly offset from their host galaxies. Fibres not used for the GW event will in parallel contribute to other science cases, minimizing disruption to the other surveys despite the ToO observing mode.

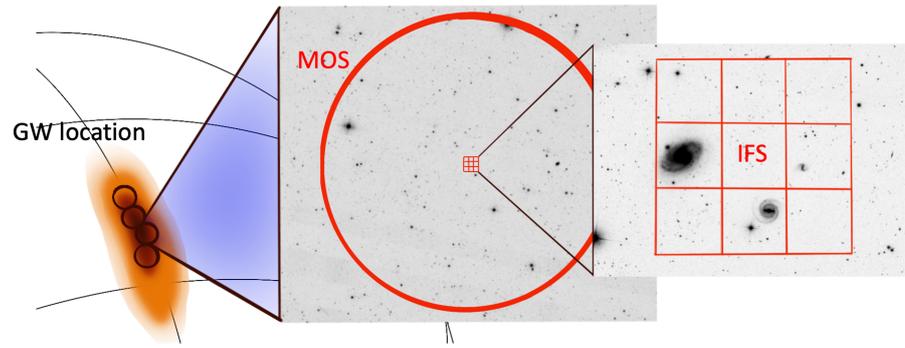

*Figure 12. WST will be superbly suited for the detection and characterization of electromagnetic counterparts to Gravitation Wave events detected by the Einstein Telescope. The IFS will be placed near the most likely host galaxy (right panel), while the wide-field MOS is required to target the large number of alerts that will be issued by Rubin LSST within the sky localization region (middle panel). Credit: S. Bisero, R.I. Anderson, and S. Vergani.*

We have summarised here just a few examples of the transformative science that WST will unlock. The Science White Paper [3] describes how WST will be able to tackle a very broad range of science areas, including: constraining the mass of neutrinos; probing primordial perturbations and testing inflationary models; studying the flow of gas in and out of galaxies and its relation with environment; understanding the growth of mass into supermassive black-holes and their impact on the galaxies hosting them; dissecting the baryonic and dark matter properties of dwarf galaxies, the most common type of galaxies found in the Universe today; studying with unprecedented detail the star-formation processes in massive and dense environments; determining precise elemental abundances and chromospheric activity in large sample of stars hosting planets; gaining insights on the formation and evolution of star clusters; unveiling the extreme physics at play during the destruction of a star approaching a massive black-hole; and tracing the chemical elements in the coma of comets transiting the Solar System. Overall, the acquisition of large and consistent samples of sources will allow the scientific community to make precise statistical comparisons of physical properties, test physical models, and potentially detect systematic anomalies.

***Synergies:*** As shown by the examples mentioned above, WST has the potential to make a significant scientific impact on its own. However, WST synergy with existing or forthcoming large ground-based and space-based facilities will greatly enhance their scientific capabilities. Thanks to its unique capabilities, WST will be an unparalleled complementary facility for Euclid, Roman, LSST/Rubin, *Gaia*, SKAO, Einstein Telescope (ET) and CTA. WST will also be a feeder facility for ESO's ELT, providing important rare sources that will be discovered by these large surveys.

For example, the LSST survey on the Vera Rubin Observatory will conduct a survey of 20,000 deg$^2$ of the sky in six broad photometric bands. The survey depth of *r*~27.5 will be achieved at the end of the 10-year survey. WST is the only facility that will have the capability to follow up such a wide-field deep survey, and its southern location makes it ideally suited for this task. The Square Kilometre Array (SKA) is a new-generation radio telescope that is set to revolutionise our view of the radio sky, performing sky surveys with broad frequency coverage (from 50 MHz up to 15 GHz), and a spatial resolution reaching 0.04 arcsec at the highest frequencies. With a 2-year integration time for an all-sky survey at 1 GHz, SKA is anticipated to achieve a 3 µJy rms sensitivity, detecting approximately 4 galaxies per arcmin$^2$ and over 0.5 billion radio sources. Only WST will be able to complete these radio surveys with the crucial optical redshift information needed for full scientific exploitation of the SKA data.

Noticeably, WST will not just be following up other facilities—its surveys will also provide a wealth of sources for other facilities to follow up in turn, for example by using the higher spatial or spectral resolution of the ELT instrumentation (e.g., MICADO, HARMONI, ANDES) or by studying them at other wavelengths (e.g., ALMA, METIS).

## 4. SYSTEM DESIGN

For WST to deliver the exceptional quality and quantity of science data demanded by the astronomy community, we believe the telescope and instruments must be designed as an integrated system with the enclosure and surrounding infrastructure, ensuring efficient operation in an environmentally sustainable manner. Traditional 'general facilities' such as the VLT, Keck, ELT etc were designed so their telescopes have a 'flexible' interface at one or more focal planes, thereby enabling an ever-evolving range of instruments to be deployed to suit different science cases and the flexible scheduling of observations. WST, as a survey facility, will be developed with a completely different design philosophy.

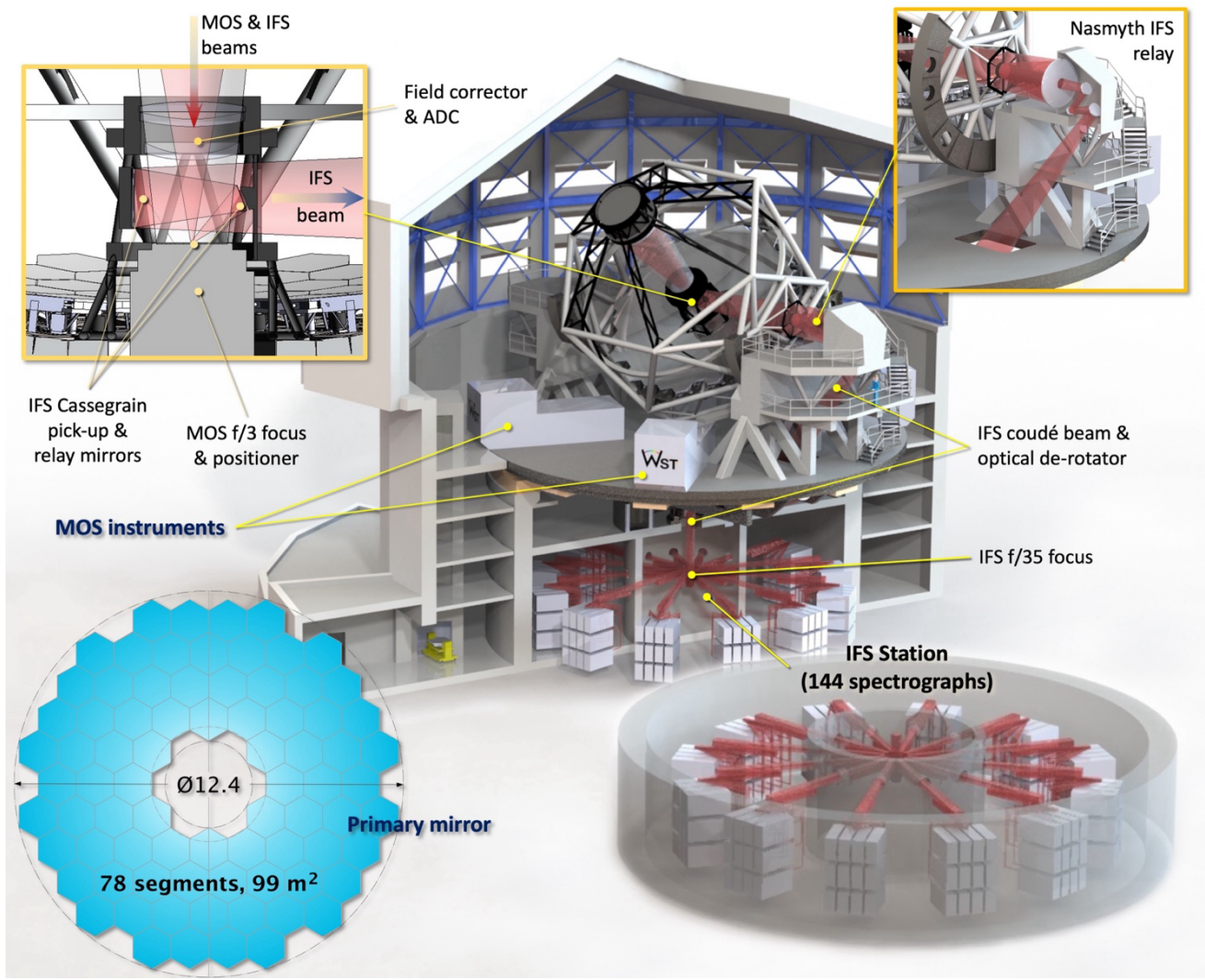

*Figure 13 – Current Facility design. The MOS spectrographs are located on the azimuth floor, for minimal fibre length. The gravity-invariant IFS station is located in the pier, below the optical de-rotator. IFS sub-field extraction (3 x 3 arcmin$^2$ out of the 13 arcmin diameter patrol field) is performed at the optical exit of the Nasmyth relay.*

The system (Figure 13) is an alt-azimuthal, 12.4-m telescope which provides two concentric, simultaneously available fields – a 3.1 sq. degree f/3 one (MOS) at a corrected Cassegrain focus located above the primary mirror, and a 13 arcmin diameter f/35 one (IFS) in a Nasmyth configuration. Within the latter, any 3 x 3 arcmin$^2$ sub-field can be selected and propagated through a Coudé relay, down to the gravity-stable IFS station located in the telescope pier. All optics are within commercially available dimensions. The volume of the fibre-fed MOS instruments are located on the azimuth floor, for minimal fibre length. The telescope and instruments designs are detailed in [5] and [6], respectively.

## 5. SITE SELECTION

We have performed an initial exploration of potential locations in or around the Paranal Armazones region in Chile (Figure 14) and have identified the *La Chira* peak as a promising candidate. This site stands out due to its strategic placement within the ESO land concession, ensuring convenient accessibility. Moreover, its distance from both the VLT and ELT (respectively 13 and 18 km) safeguards against potential adaptive optics laser collisions. Notably, the *La Chira* peak offers the added advantage of proximity to the VLT/ELT base camp and solar power plant. This advantageous location shortens the access road and minimises the environmental impact by reducing $CO_2$ emissions. The concept study will make use of both archival and purpose-built turbulence measurements at *La Chira* and other candidate sites. The preferred site will be documented and its characteristics (e.g. turbulence, wind statistics) will be used in the telescope and instruments designs trade-offs.

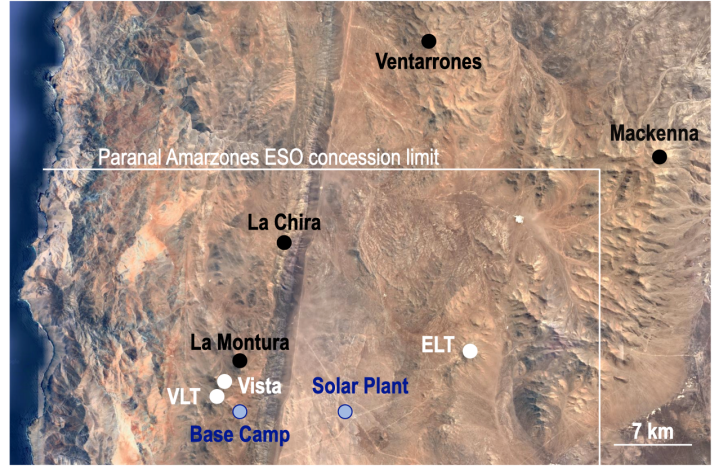

*Figure 14: Potential sites for WST in or near the Paranal Armazones area (in black).*

## 6. CONSORTIUM

Designing such an ambitious facility, including so many different subsystems and features, is both a demanding and challenging task. As mentioned, we have raised significant interest within the community and brought together a large consortium of 26 research institutes or universities, including scientists from ESO, spread over 10 different countries (9 countries in Europe plus Australia).

The consortium gathers all the required expertise to perform the study with close collaboration between consortium members for most of the work-packages. For critical systems we will have multiple experts working in parallel on alternative designs during the tradeoff phase (e.g., EPFL, AIP, MACQ and UKRI for the MOS positioner task). This is a strong asset to find the best design. In addition to having significant experience in running spectroscopic surveys, the whole consortium has an impressive record (past/present) in instrumentation and telescope projects. Together we have built (or are studying and building) 9 optical telescopes, 8 MOS and 10 IFS.

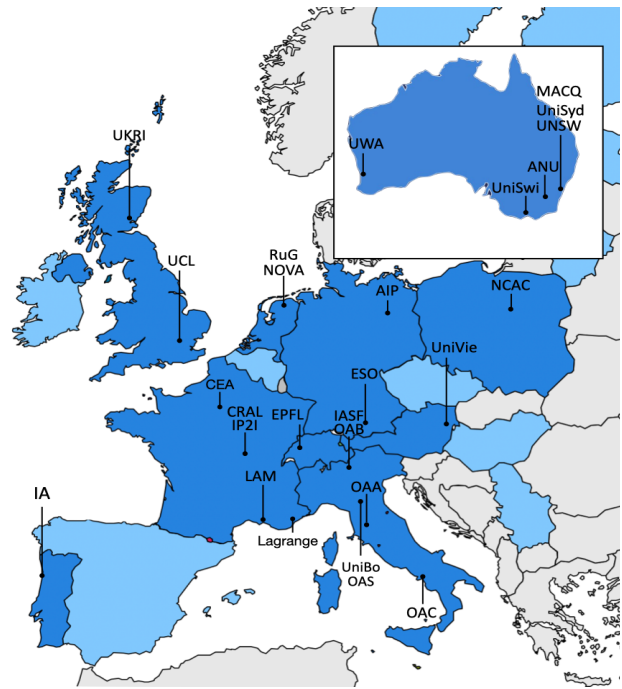

*Figure 15: The distribution of participant's institutes (dark blue) and science team countries (light blue) in Europe and Australia.*

## 7. CONCLUSIONS

We have described the wide-field spectroscopic survey telescope (WST) project. A 12-metre wide-field spectroscopic survey telescope with simultaneous operation of a large field-of-view (3 sq. degree), high-multiplex (20,000) multi-object spectrograph (MOS) and a giant 3×3 arcmin integral field spectrograph (IFS). We have shown how WST can address outstanding scientific questions in the areas of cosmology; galaxy assembly, evolution, and enrichment, including our own Milky Way; the origin of stars and planets; and time domain and multi-messenger astrophysics.

We will propose WST as the next major European Southern Observatory (ESO) project after ELT. In the long run, the realisation and operation of WST at ESO would fully leverage its well-proven capabilities and experience in developing and operating large research infrastructures, while efficiently exploiting its already strong connections to the European industrial and scientific communities. It is timely because the workforce and financial resources that ESO currently dedicates to the 39m ELT construction will gradually become available as we approach and pass ELT's first light, expected in 2028. This strategy also fits within the framework of ESO plan to launch a **"Call for Ideas"** in the coming year to identify potential future major projects, and ultimately to select one to follow the ELT.

Acknowledgments: We acknowledge the fundamental contributions from the WST science team. Individual science cases reported here have been developed by team members and their individual contributions are listed in Mainieri et al 2023. RB, PD, LT thanks CNRS INSU CSAA and ANR-AA-MRSE-2023 for their support. PS, SR, BG acknowledge support from INAF - Astrofisica Fondamentale 2023 Large Grant "The WST". Australia is currently a strategic partner in the European Southern Observatory, supported by the Australian government's 2017-18 Budget measure *Maintaining Australia's Optical Astronomy Capability*."